\documentclass[%
reprint,
superscriptaddress,
nofootinbib,
amsfonts,amsmath,amssymb,
prb,
longbibliography
]{revtex4-1}
\usepackage[utf8]{inputenc}
\usepackage{amsmath,amsthm}
\usepackage{dcolumn}
\usepackage{bm}
\usepackage[colorlinks, allcolors=blue]{hyperref}
\usepackage[pdftex]{graphicx}
\usepackage{braket}
\usepackage{color}
\usepackage{natbib}
\usepackage[usenames,dvipsnames,svgnames,table]{xcolor}

\begin{document}

\title{Tunneling density of states in a Y junction of Tomonaga-Luttinger liquid wires: A density matrix renormalization group study}

\author{Monalisa Singh Roy}
\email{monalisa12i@bose.res.in}
\affiliation{S. N. Bose National Centre for Basic Sciences, Block - JD, Sector - III, Salt Lake, Kolkata - 700 106, India}

\author{Manoranjan Kumar\textsuperscript{\textsection}}
\email{manoranjan.kumar@bose.res.in}
\affiliation{S. N. Bose National Centre for Basic Sciences, Block - JD, Sector - III, Salt Lake, Kolkata - 700 106, India}

\author{Sourin Das\textsuperscript{\textsection}}
\email{sourin@iiserkol.ac.in}
\affiliation{Indian Institute of Science Education And Research Kolkata, Mohanpur, Nadia - 741 246, West Bengal, India}

\thanks{These authors contributed equally to this work.\\}  \label{Footnote}

\date{July 17, 2020}

\begin{abstract} 

It is well known that the pristine bulk of an interacting one-dimensional system in Tomonaga-Luttinger liquid (TLL) phase shows power law suppression of quasi-particle tunneling amplitude for all values of TLL parameter $g$, in the zero energy limit. We perform a density matrix renormalization group (DMRG) study of a fully symmetric Y junction of TLL wires and observe an anomalous enhancement of the tunneling density of states (TDOS) in the vicinity of the junction for both (a) interacting bosons case and (b) interacting fermions case, when $g>1$. We also observe suppression of TDOS for $g<1$ for both bosonic and fermionic cases.  We find that the TDOS enhancements follow different power laws for  bosonic and fermionic cases which suggests that these represent distinct fixed points, owing to statistical correlations which play an important role at the Y junction. Analysis of static conductance for the junction indicates that the fixed point for $1<g<3$ resembles the mysterious $M$ fixed point of Y junction predicted by Oshikawa, Chamon, and Affleck [J. Stat. Mech. P02008 (2006)]. {We also show that the TDOS enhancement spans over a length scale of $\propto \omega^{-1}$ from the junction, for $1<g<3$.}

\end{abstract}


\maketitle


\section{Introduction} \label{INTRODUCTION}
The technological advances at sub-micron scales have enabled fabrication of one-dimensional (1D) wires and their junction with high precision~\cite{1997Tans,1999_Bockrath,2000_Auslaender,2002_Auslaender,2003_Shiraishi}. In a confined quasi-1D geometry, effect of inter-electronic repulsion is omnipresent, and the weakest of interactions could drive the system to the Tomonaga-Luttinger liquid (TLL) phase in the low energy limit.~\cite{2004_Giamarchi,1992_Goldenfeld,1998_vonDelft} The TLL phases~\cite{1981_Haldane,1981b_Haldane} of 1D electronic quantum systems have been of sustained interest to condensed matter physicists due to their non-Fermi liquid  behavior.~\cite{1950_Tomonaga,1963_Luttinger,1981_Haldane,2004_Schoenhammer,2012_Schoenhammer,2017_Caux} The power law decay of the bulk electronic density of states (DOS), $\rho (\epsilon) \sim |\epsilon - \epsilon_F|^{\alpha}$ ($\epsilon_F$ being the Fermi energy) is a well known signature of TLL wires, where the value of $\alpha$ depends on the system parameters. Here $\alpha>0$  indicates the fact that the DOS goes to zero as the energy approaches the Fermi energy which is an effect induced purely due to inter-particle interaction. 

An early study of tunneling into a TLL wire was reported by Oreg and Finkelstein~\cite{1996_Oreg} and since then there have been several works reported on the topic.~\cite{1997_Nazarov,2009_Agarwal,2010_Aristov,2013_Jeckelmann,2015_Mardanya,2018_Latief,2019_Vu} Amongst these, Jeckelmann in Ref.~[\onlinecite{2013_Jeckelmann}] applied dynamical density matrix renormalization group (DMRG) method to a 1D spinless fermion (SF) chain with nearest neighbor interaction. They confirmed that the bulk DOS shows a power law suppression as $\epsilon \rightarrow \epsilon_F$ in the gapless phase, as is expected from the TLL theory. They also confirmed  that the tunneling density of states (TDOS) shows an enhancement (suppression) as $\epsilon \rightarrow \epsilon_F$ at the boundary of the SF chain for attractive (repulsive) inter-particle density-density interaction, which is consistent with the predictions of TLL theory~\cite{1997_Fisher}. 

An interesting variant of the two-terminal TLL wire set up is the junction of three or more TLL wires. Such multi-wire junction of TLL wires presents a quantum impurity problem which is distinct from an isolated quantum impurity embedded in the bulk of a pristine TLL owing to its much richer fixed point structure. In recent times, junctions of TLL wires have gained much interest, especially the three-wire junction (Y junction) which is the simplest non-trivial junction of 1D TLL wires. This structure can be recognized as a basic constituent of future quantum circuits and has already been explored experimentally.~\cite{1999_Yao,1999_Li,2001_Egger,2004_Biro,2009_Subhramannia,2015_Ding,2018_Sharma,2018_Mosallanejad} The first theoretical work on this topic was reported by Nayak et al., where they used bosonization and boundary conformal field theory techniques to obtain  fixed point conductance of the Y junction hosting a resonant level.~\cite{1999_Nayak} Since then the studies on the topic has predominantly focused on finding various interesting fixed points and analyzing the spectral properties of the system using bosonization, weak interaction renormalization group (WIRG) or functional renormalization group(fRG).~\cite{1999_Nayak,2002_Lal,2003_Egger,2004_Das,2005_Barnabe,2005b_Barnabe,2006_Kakashvili,2006_Das,2008_Tokuno,2009_Watcher,2009_Agarwal,2010_Rahmani,2011_Aristov,2012_Rahmani,2012_Aristov,2013_Aristov,2015_Mardanya,2003_Chamon,2006_Oshikawa,2000_Meden,2009_Bellazzini,2012_Calabrese} {In particular, an exhaustive study of various fixed points of a Y junction enclosing a central flux ($\phi$), and their corresponding conductances was reported by Oshikawa et al. using bosonization and boundary conformal field theory techniques. They conjectured the existence of a stable  ``mysterious'' $M$ fixed point ($\phi=0$ condition) in the attractive interaction regime $1<g<3$.~\cite{2006_Oshikawa} However, they also concluded that the conformally invariant boundary condition describing this fixed point could not be identified and it remains an open problem. Later Rahmani et al. developed a method to evaluate the conductance of junction of  multiple TLL wires  using static ground state (gs) correlations and applied it to the $M$ fixed point where the ground state was obtained numerically.~\cite{2012_Rahmani}}  

Studies of TDOS using bosonization technique for a Y junction of TLL wires was reported by Agarwal et al. in Ref.~[\onlinecite{2009_Agarwal}] and a collection of fixed points were identified which showed enhancement of TDOS in the zero frequency limit. This effect was attributed to an Andreev-like reflection off the junction. This study was later extended to include spin degrees of freedom in Ref.~[\onlinecite{2015_Mardanya}]. The ground state properties of Y junctions have also been explored using DMRG techniques.~\cite{2006_Guo,2016_Kumar,2019_Buccheri}. However it should be noted that a numerical study using dynamical DMRG techniques focused on evaluation of TDOS for Y junction is presently lacking in literature, and is the primary focus of the present work. 

This paper starts by considering a Y junction of  spin$-1/2$ chains with nearest neighbor anisotropic (XXZ) Heisenberg type interaction. This model can be exactly mapped on to a hard-core boson (HB) model with nearest neighbor interaction. We perform a DMRG study of Y junction for the XXZ model and the corresponding SF model.  We  use the correction vector approach to calculate the local contribution to the TDOS of the system.~\cite{1989_Soos,1995_Ramasesha,1999_Pati,2002_Jeckelmann} We first study the Y junction of SF chains and draw a comparison with the existing studies of 1D SF chains and report enhancement of TDOS in $g>1$ limit. Thereafter, we shift our focus to the XXZ Y junction and verify the existence of enhancement in TDOS near the junction in $g>1$ limit. {We also demonstrate that the enhancement of TDOS is related to the $M$ fixed point. It should be noted that the evaluation of TDOS requires dynamical correlations functions as input. The previous study by Rahmani et al.~\cite{2012_Rahmani} used time-independent DMRG to calculate the static ground state correlations, while we evaluate the dynamical correlations using dynamical DMRG techniques for the $M$ fixed point, hence enriching the existing understanding of this analytically unsolvable problem of $M$ fixed point. Next, we explore the finite size effect on the TDOS spectra in the enhancement regime, and comment on the length scale of the observed TDOS enhancement near the junction.}

This paper is organized in four sections. The motivation and existing studies related to our problem have been introduced in Sec.~\ref{INTRODUCTION}. The model and numerical techniques are described in detail in Sec.~\ref{MODEL AND NUMERICAL TECHNIQUE}. The calculation of TDOS for the system using the correction vector method has been explained there. The results are described in Sec.~\ref{RESULTS AND DISCUSSIONS}. We have concluded by summarizing  our findings in Sec.~\ref{SUMMARY} .

\section{Model and Numerical Techniques} \label{MODEL AND NUMERICAL TECHNIQUE}
\begin{figure}[t] 
\includegraphics[width=0.6\linewidth]{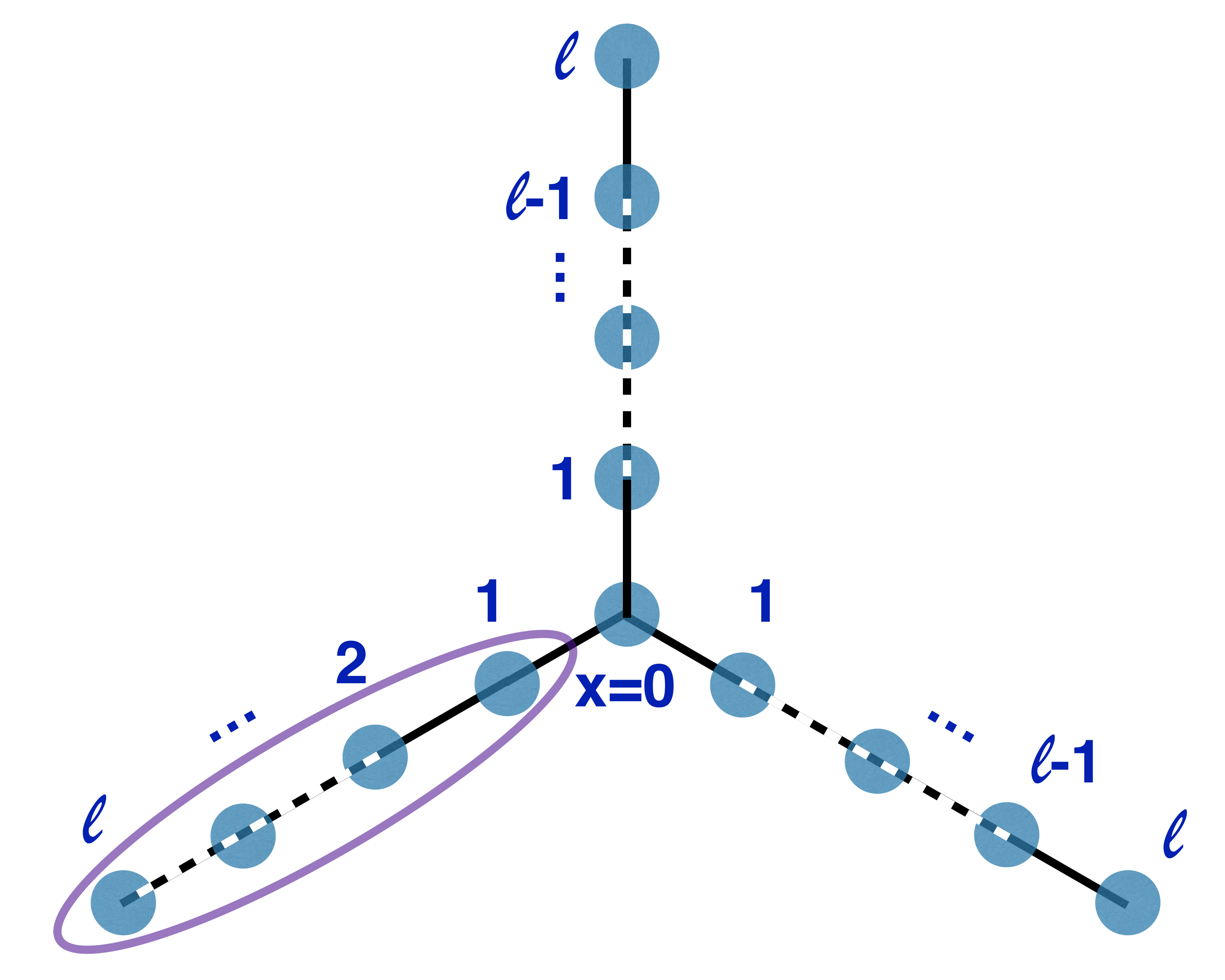}
\caption{(Color online) Schematic of Y junction of size $N=3\ell+1$ sites, formed by three 1D TLL wire arms of length $\ell$ each (encircled), joined at an additional central site, at $x=0$. In our convention, the labeling of the spin sites start from the junction, as illustrated in the figure.} \label{fig1: Yjn}
\end{figure}

We consider a Y junction of $N=3\ell+1$ sites, constituted by three 1D TLL wires of $\ell$ sites each, connected at a common central site labeled $x=0$, as shown in schematic Fig.~\ref{fig1: Yjn}. 
Our goal is to study both the bosonic and the fermionic Y junction models. We start by considering a Y junction of three spin-$1/2$ chains, where spins are interacting with their nearest neighbors only through an anisotropic (XXZ) Heisenberg like interaction.
The model Hamiltonian for the system is given by

\begin{align} \nonumber \label{eq.Hamiltonian.Jz}
{\cal H}=&\sum^{\ell-1,3}_{x=1,k=1} \left[ \dfrac{J}{2} (S_{x,k}^{+} S_{x+1,k}^{-} + h.c.) + J^z S_{x,k}^{z} S_{x+1,k}^{z} \right] \\ \nonumber
+& \sum_{k=1}^{3} \left[ \dfrac{J}{2}(S_{0}^{+} S_{1,k}^{-} + h.c.) + J^z S_{0}^{z} S_{1,k}^{z} \right] , \\ 
\end{align} 

where $S_{x,k}^+$ $(S_{x,k}^-)$ and $S_{i,k}^z$ are the spin raising  (lowering) operator and $z-$component of local spin operator, respectively, acting at lattice site $x$ on leg $k$ of the system. $S_{0}^+$ $(S_{0}^-)$ and $S_{0}^z$ are the spin raising (lowering) operator and $z-$component of local spin operator, respectively, acting at the junction site $x=0$. The first part of the Hamiltonian represents exchange interactions in each of the three wires (labeled by $k=1,2,3$). In the present work, we consider the XXZ model Hamiltonian, therefore we have taken $J^x= J^y=J$ and the value $J=1$ has been kept fixed in all the calculations related to the XXZ Y junction, and $J^z$ is the variable parameter.

Next, we consider the Y junction of HB wires where the bosons obey only nearest neighbor inter-particle interaction, and the corresponding Hamiltonian can be written as
\begin{align} \nonumber \label{eq.Bose-Hubbard} 
{ \cal H} =& \sum^{\ell-1,3}_{x=1,k=1} \left[ -t ( b_{x,k}^{\dagger} b_{x+1,k}+ h.c.) \right. \\ \nonumber
& \left. + V n_{x,k} n_{x+1,k} + \mu \left(n_{x,k}+\frac{1}{4} \right) \right] \\ \nonumber
& + \sum_{k=1}^{3} \left[ -t(b_{0}^{\dagger} b_{1,k} + h.c.) + V n_{0} n_{1,k}  \right]  + 
\mu \left( n_{0}+\frac{1}{4} \right) , \\ 
\end{align} 

where $b_{x,k}$ $(b^{\dagger}_{x,k})$ and $n_{x,k}$ are the boson annihilation (creation) operator and the number operator, respectively, acting at lattice site $x$ of leg $k$. $b_{0}$ $(b^{\dagger}_{0})$ and $n_{0}$ are the boson annihilation (creation) operator and occupation number operator, respectively, acting at the junction site $x=0$. In the HB limit, the maximum occupation number of the each site is $1$, i.e., each site possesses two degrees of freedom, similar to the spin$-1/2$ system. The Hamiltonian in Eq.~\eqref{eq.Hamiltonian.Jz} can be exactly mapped to this bosonic Hamiltonian in Eq.~\eqref{eq.Bose-Hubbard}, through the transformation $t= -J/2$, $V=J^z$ and $\mu=J^z$, where $t$, $V$, and $\mu$ are the transfer integral, density-density interaction strength between neighboring sites, and the chemical potential strength of the system, respectively~\cite{2011_Sachdev}. Since there is a one-to-one mapping between the HB and XXZ spin$-1/2$ and the whole energy spectrum is same, we solve only the XXZ model and refer to it as the bosonic Y junction.

Finally, we consider the SF model on the Y junction geometry, where the fermions obey only nearest neighbor inter-particle interaction, and the corresponding Hamiltonian can be written as
\begin{align}  \nonumber
\label{eq.spinless_fermions}
{\cal H}=&\sum^{\ell -1,3}_{x=1,k=1} \left[ -t (c_{x,k}^{\dagger} c_{x+1,k}+ h.c.) \right. \\ \nonumber
& \left. + V n_{x,k} n_{x+1,k} + \mu \left( n_{x,k}+\frac{1}{4} \right)  \right] \\ \nonumber 
& +\sum_{k=1}^{2} -t(c_{0}^{\dagger} c_{1,k} + h.c.) -t^{\prime} (c_{0}^{\dagger} c_{1,3} + h.c.) \\ \nonumber 
& +\sum_{k=1}^{3} V n_{0} n_{1,k} + \mu \left( n_{0}+\frac{1}{4} \right) , \\ 
\end{align} 
where $c_{x,k}$ $(c^{\dagger}_{x,k})$ and $n_{x,k}$ are the fermion annihilation (creation) operator and the occupation number operator, respectively, acting at site $x$ of leg $k$. $c_{0}$ $(c^{\dagger}_{0})$ and $n_{0}$ are the fermion annihilation (creation) operator and number operator, respectively, acting at the junction site $x=0$. The model Hamiltonian in Eq.~\eqref{eq.Hamiltonian.Jz} can be mapped to this fermionic model Hamiltonian in Eq.~\eqref{eq.spinless_fermions} using Jordan-Wigner (JW) transformation~\cite{1934_Jordan} through the parameter transformations as: hopping integral $t=-J/2$, electron-electron interaction $V=J^z$ and the chemical potential $\mu=J^z$. $t^{\prime}$ in Eq.~\eqref{eq.spinless_fermions} can be related to $t$ in Eq.~\eqref{eq.Bose-Hubbard} as $t^{\prime} =  \prod_{x=1}^{\ell} (-1)^{n_{x,2}+n_0} t$ (the site labels are shown in schematic Fig.~\ref{fig1: Yjn}). It is easily seen that Eq.~\eqref{eq.spinless_fermions} is essentially same as Eq.~\eqref{eq.Hamiltonian.Jz} and Eq.~\eqref{eq.Bose-Hubbard} for a linear 1D chain. However, for the multi-wire junction, the SF system is distinguished due to the non-trivial phase factors associated in the hopping interaction $t'$ between the junction and the third constituent wire, which accumulates the delocalized JW phase from the other two constituent wires. We refer to this SF Y junction system as the fermionic Y junction. {It should be emphasized that this excess phase in the fermionic case is not a single particle phase, rather it is a many-body phase which depends on the occupancy of fermions at the central site and the other constituent chains. When we are in the TLL phase, the electron is delocalized and hence the  occupancy of fermion at the central site is a dynamical quantity. So, the difference between the bosonic and fermionic case can be thought of as a difference of having or not having a dynamical phase factor associated with the junction site. Further, it should be noted that this extra phase which distinguishes the Y junction of bosonic chains from the Y junction of fermionic chains can not be thought of as a small difference since it can have non-trivial consequences in deciding the stable fixed point for the Y junction. This difference would also be reflected later in the TDOS power laws for both models. In continuum model of TLL these extra phase are introduced into the tunneling Hamiltonian forming the junction via Kline factors and a detailed discussion on the influence of their presence in deciding stable fixed point of Y junction can be found in Ref.~[\onlinecite{2003_Chamon}].} 
In our numerical analysis using DMRG for the fermionic Y junction, we have kept $t'=t=1$ fixed for all calculations. In this paper we study both the bosonic and fermionic Y junction models.

To correlate our lattice model parameters with the TLL parameter, we use the results from the 1D bosonic and fermionic systems. The TLL parameter $g_s$ corresponding to the exchange interaction $J^z$ of 1D spin$-1/2$ or bosonic system can be derived using Bethe ansatz (a derivation is presented in Ref.~[\onlinecite{2001_Rao}]), and is given by
\begin{equation} \label{eqn.LuttingerXXZ}
\dfrac{1}{g_s}=1+\dfrac{2}{\pi}sin^{-1}\left(\dfrac{J^z}{J}\right) .
\end{equation}
The TLL parameter $g_f$ corresponding to the inter-particle density-density interaction $V$ in the half-filled 1D fermionic model can be derived using Bethe Ansatz (a derivation is presented in Ref.~[\onlinecite{2004_Schoenhammer}]), and is given by
\begin{equation} \label{eqn.LuttingerSF}
g_f = \dfrac{\pi}{2} \hspace{1ex} \dfrac{1}{\pi - cos^{-1}\left(V/2t\right)}  .
\end{equation}
The limit $J^z=0$ $(V=0)$ corresponds to the free-particle limit, where $g_s=1$ $(g_f=1)$. The ferromagnetic $J^z<0$ (or attractive limit $V<0$) corresponds to the TLL parameter $1 < g_s (g_f) <\infty$, and the antiferromagnetic $J^z>0$ (or repulsive limit $V>0$) corresponds to $0 < g_s (g_f) < 1$. We study the TDOS in the bosonic and the fermionic Y junction systems in both the $0 < g_s (g_f) <1$ and $1 < g_s (g_f) < 3$ limits to identify the enhancement and suppression regimes.

Since all the model Hamiltonians considered on the Y junction geometry in Eq.~\eqref{eq.Hamiltonian.Jz},~\eqref{eq.Bose-Hubbard}, and~\eqref{eq.spinless_fermions} contain many-body interaction terms, hence, the degrees of freedom in the system increases exponentially with the system size $N$. Therefore, the exact diagonalization (ED) technique is used for system sizes up to $N \sim 28$,  and DMRG technique is used for larger system sizes, up to $N=610$. DMRG is a state-of-art numerical technique based on the systematic truncation of irrelevant degrees of freedom, and renormalization of the system observables with the reduced density matrix wavefuncion.~\cite{1992_White,2005_Schollwock} For accurate calculations we have used the modified DMRG algorithm especially designed for Y junction, which renders the accuracy of these calculations comparable to that for linear 1D chains.~\cite{2016_Kumar} To maintain a reliable accuracy in the calculations, eigenvectors corresponding to $\sim 200$ largest eigenvalues of the density matrix are retained in each DMRG sweep. The truncation error of the density matrix eigenvalues is less than $10^{-12}$. For better accuracy, we perform finite DMRG upto $10$ sweeps, and the total error in the ground state is less than $0.01\%$.

In this paper, study of TDOS is our main focus. {TDOS is equivalent to locally injecting a magnon into the ground state of the system, which can access all the excited states with a finite transition probability determined by the non-zero transition matrix elements between the ground state and the respective excited state. Thus,} the TDOS for a system gives information about the low lying excitations in the system, and can be defined as,
\begin{align} \nonumber
\label{eq.xTDOS}
\rho_{x} (\omega) = & \int_0^\infty e^{-(i\omega t-\eta)} dt | \langle \psi_0 |A_x(t) A_x^{\dagger}(t)| \psi_{0} \rangle \\  \nonumber
=& \int_0^\infty \sum_n  e^{-(i\omega t-\eta)} dt | \langle \psi_0 |A_x(0) e^{-iHt} | \psi_{n}\rangle \\  \nonumber
  & \times \langle \psi_n |e^{iHt} A_x^{\dagger}(0) | \psi_{0} \rangle \\ \nonumber
\propto  & \hspace{1ex} \text{\normalfont Im} \left[ \sum_{n} \dfrac{| \langle \psi_n | A_{x}^{\dagger} | \psi_{0} \rangle |^{2}} {E_{n}- (E_{0} + \omega) +i \eta} \right]  , \\ 
\end{align}
where $\vert \psi_{0} \rangle$ and $E_0$  represent the ground state wavefunction and energy, respectively. $\vert \psi_n \rangle$ and $E_n$ represent the eigenvector and eigenvalue corresponding to the $n^{th}$ eigenstate of the system, respectively. $A_x^{\dagger}$ represents the spin raising operator ($S^+_x$), the boson creation operator ($b_x^{\dagger}$), or the fermion creation operator ($c_x^{\dagger}$), acting at site $x$ in Eq.~\eqref{eq.xTDOS}. The spatial numbering in the Y junction system is shown in Fig.~\ref{fig1: Yjn}. The broadening factor $\eta$ used in the calculation of TDOS in Eq.~\eqref{eq.xTDOS} is generally proportional to the lifetime of quasi-particles. {It helps in avoiding the unphysical divergence in $\rho_{x}(\omega)$ at the Fermi energy and it induces a Lorentzian behavior in $\rho_{x}(\omega)$ near resonance frequency $\omega_p=E_{n}-E_{0}$. This does not change the physics of the problem, and to extract the power law exponents ($\alpha$) of $\rho_{x}(\omega)$ as a function of $\omega$, we fit $\rho_{x}(\omega)$ with power law function for $\omega > \eta$.} $\eta=0.20$ has been kept fixed throughout all the calculations. Both $\omega$ and $\eta$ have been described everywhere in units of $t$. We use the TDOS correction vector technique to calculate the TDOS, which is a state-of-art numerical technique for dynamical calculations~\cite{1989_Soos,1995_Ramasesha,1999_Pati,2002_Jeckelmann}.

\section{Results and Discussions} \label{RESULTS AND DISCUSSIONS}

In this paper, we present TDOS behavior of both the bosonic and the fermionic Y junction systems. We find that the TDOS in the proximity of junction (including the junction site) shows enhancement in the attractive interaction limit and suppression in the repulsive interaction limit, for both the bosonic and the fermionic Y junction models. These results have also been complemented by the static conductance calculations which lead to identification of the fixed points responsible for the observed enhancement or suppression of the TDOS. In particular, we show that the fixed point corresponding to the enhancement in the fermionic Y junction model belongs to the $M$ fixed point earlier predicted by Oshikawa et al. in Ref.~[\onlinecite{2006_Oshikawa}]. Though we observe similar signatures in the static current-current conductance for both the bosonic and the fermionic Y junction models, the power law exponents for the TDOS near the junction for both systems are distinct, which can be attributed to the exchange statistics of the respective particles\--- bosons in the bosonic Y junction model, and fermions in the fermionic Y junction model. We note here that for a two-wire junction, the effect of statistics of the particles generally is not reflected in the TDOS spectrum, owing to cancellation of the statistical phase in 1D linear chain. 
In the last subsection \ref{sub.Cutoff}, we analyze the length scale of the TDOS enhancement and demonstrate that it is highly localized near the junction. Before explaining the TDOS results, let us first revisit the ground state properties of both models on the Y junction geometry. 
 
The ground state of fermionic Y junction systems for odd $N$-sized system (even $\ell$-sized constituent chain lengths) contains $\rho=N/2+1$  fermions, for an isotropic interaction $t=V$; whereas for the spin$-1/2$ Y junction (bosonic Y junction) model, the ground state lies in $S^z=1/2$ manifold at $J=J^z$. For even $N$ system size (odd $\ell$) at the isotropic interaction limit $t=V$, the ground state of the spin$-1/2$ system has three spin$-1/2$ up spins delocalized at the edge of each leg and a down spin delocalized near junction sites; however, overall the ground state of the system is a triplet state. In the anisotropic limit $J_z/J<1$ ($V/t<1$), the ground state of the spin$-1/2$ or bosonic Y junction system (fermionic Y junction system) lies in $S^z=0$ ($\rho=N/2$) sector. ~\cite{2016_Kumar}.

\subsection{Tunneling density of states (TDOS)} \label{sub.TDOS}

\begin{figure}
\includegraphics[width=0.48\textwidth]{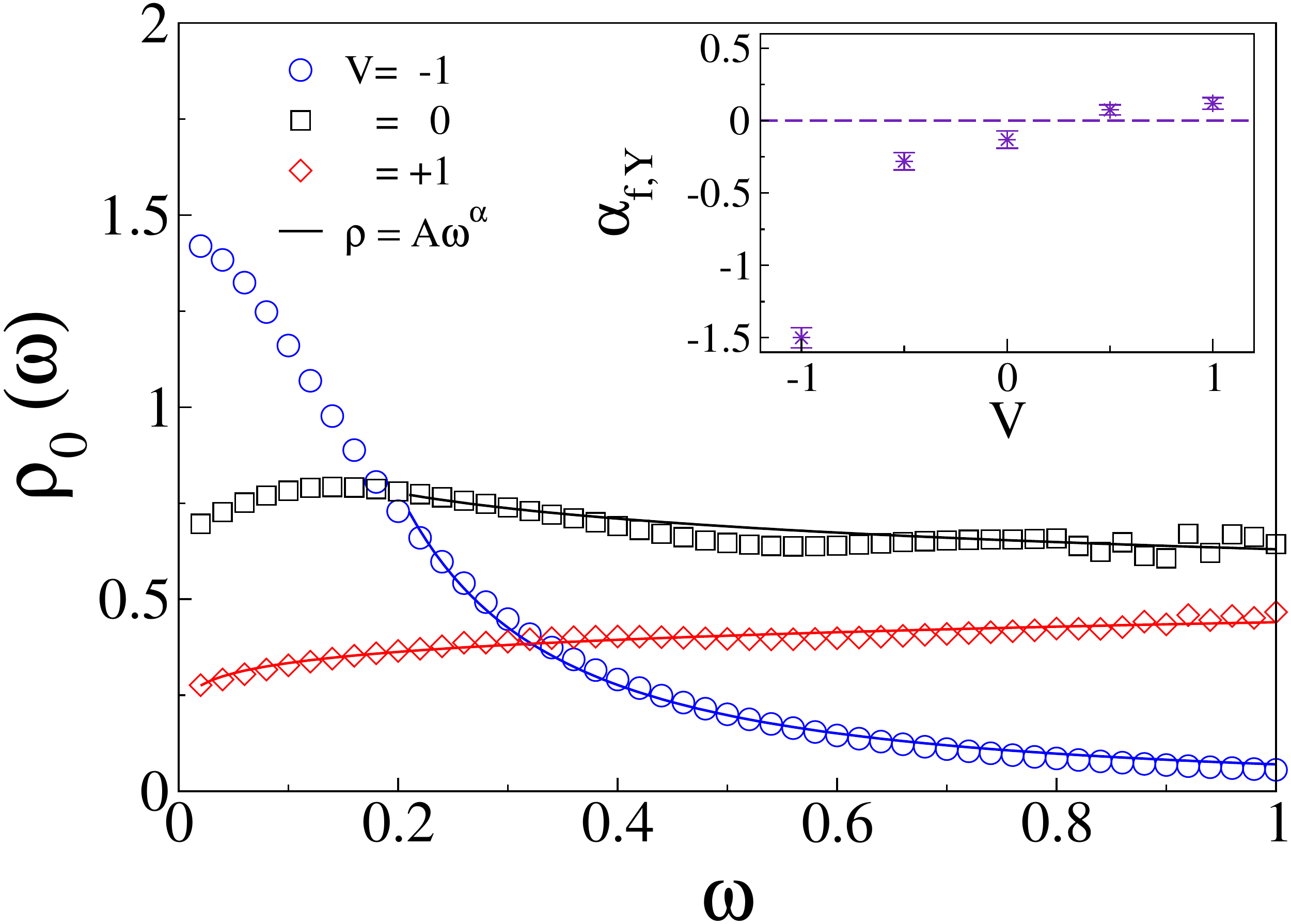} 
\caption{(Color online) The TDOS spectrum of junction ($x=0$) for the fermionic Y junction system $\rho_0 (\omega)$ as a function of frequency $\omega$, at $V=-1,0,+1$ {(or equivalently, $g_f=3/2,1,3/4$ from Eq.~\eqref{eqn.LuttingerSF})}, for a finite system size $N=310$, with broadening factor $\eta=0.20$. The solid lines show fitting of $\rho_0(\omega)$ with power law function of the form $\rho=A\omega^{\alpha}$. The fitting parameters $(A,\alpha)$ corresponding to $V=-1,0$ and $+1$ are $(0.070,-1.50)$, $(0.62,-0.13)$ and $(0.44,0.12)$, respectively. Inset: Power law exponents $\alpha_{\text{\normalfont f, Y}}$ {with error bars}, for different $V$ ($t=1$ is kept fixed).} \label{fig2}
\end{figure} 

The TDOS spectrum for 1D TLL wires has been extensively studied in literature, where the bosonic and the fermionic model spectra are indistinguishable. However for quasi-1D or multi-wire junctions, such as a Y junction, difference in TDOS spectrum is expected between the bosonic and the fermionic Y junction systems because of non-trivial many-body phase factors involved in the fermionic Y junction model, any well defined analytic study of which is lacking in literature. As the Y junction systems are well known for their unique behavior of DOS near the junction~\cite{2006_Oshikawa}, here we study the TDOS of this system near the junction for both the bosonic and the fermionic Y junction models. 
Since the TDOS of the 1D SF model has been extensively studied~\cite{2013_Jeckelmann}, therefore, we first recapitulate the TDOS results of the 1D SF system, and then compare it with that of the Y junction system. The power law exponent $\alpha$ corresponding to the TDOS of the bulk or mid-chain $\alpha_{\text{\normalfont bulk}}$, and TDOS of the boundary or open end $\alpha_{\text{\normalfont end}}$ of the interacting 1D SF chain are given by
\begin{align} \nonumber
\alpha_{\text{\normalfont bulk}} & = \dfrac{(g_f-1)^2}{2g_f} \hspace{10ex} \text{\normalfont and,}  \\ \nonumber
\alpha_{\text{\normalfont end}}  & = \dfrac{1}{g_f} -1 , \label{eqn.powerlaw1D} \\ 
\end{align}
where $g_f$ is the Luttinger parameter, as defined in Eq.~\eqref{eqn.LuttingerSF}

\begin{figure}
\centering
\includegraphics[width=0.48\textwidth]{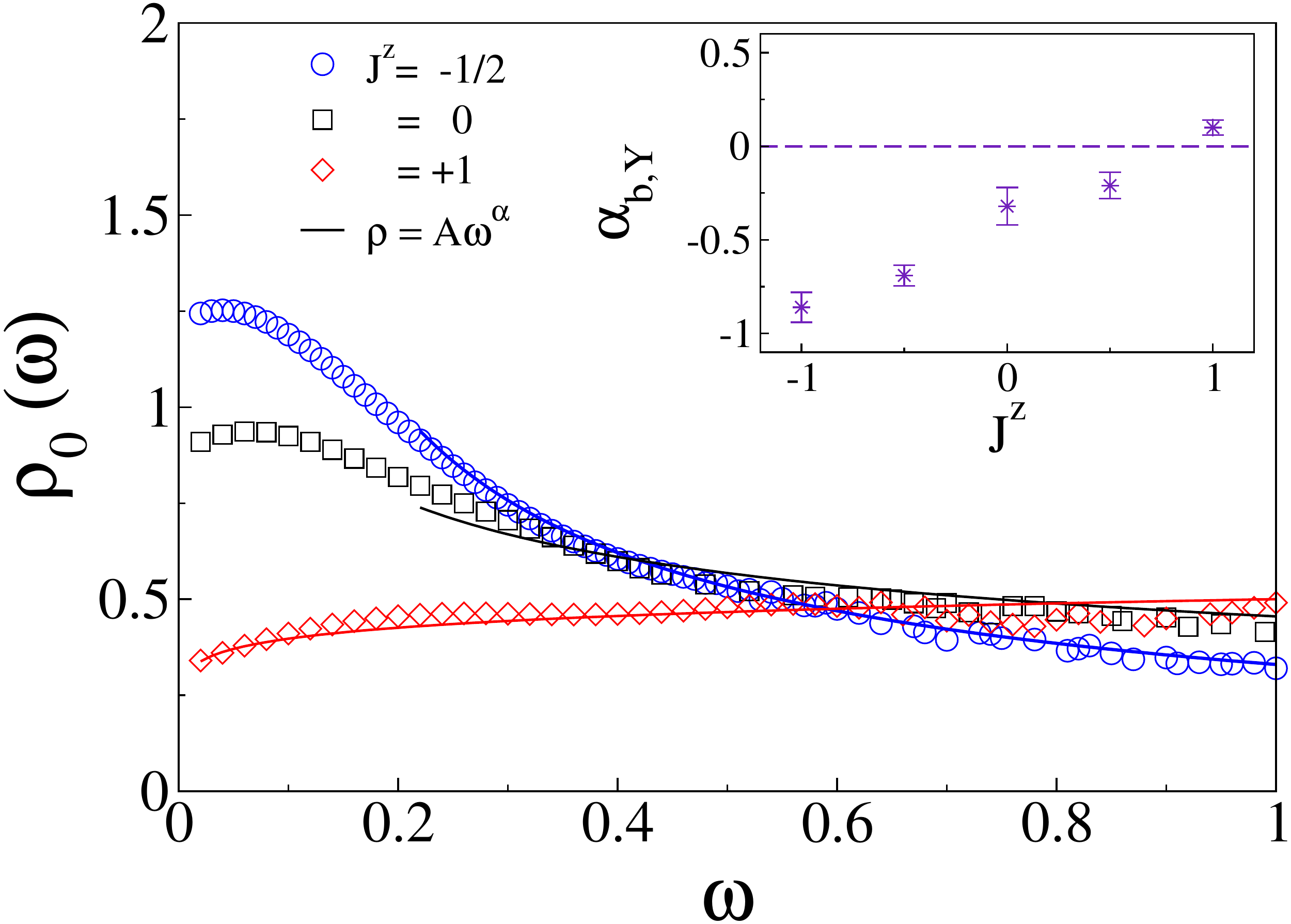} 
	\caption{(Color online) The TDOS spectrum of junction ($x=0$) for the bosonic Y junction system, $\rho_0 (\omega)$ as a function of frequency $\omega$, at $J^z=-1/2,0,+1$ (or equivalently, $g_s=3/2,1,1/2$ from Eq.~\eqref{eqn.LuttingerXXZ}), for a finite system size $N=406$, with broadening factor $\eta=0.20$. The solid lines show fitting of $\rho_0(\omega)$ with power law function of the form $\rho=A\omega^{\alpha}$. The fitting parameters $(A,\alpha)$ corresponding to $J^z=-1/2,0$ and $+1$ are $(0.33,-0.69)$, $(0.46,-0.32)$ and $(0.50,0.10)$, respectively. Inset: Power law exponents $\alpha_{\text{\normalfont b, Y}}$ {with error bars}, for different $J^z$ ($J=1$ is kept fixed). } \label{fig3}
\end{figure} 

To compare the power law exponent $\alpha_{\text{\normalfont bulk}}$ and $\alpha_{\text{\normalfont end}}$ obtained for the 1D SF chain with that obtained for the fermionic Y junction system ($\alpha_{\text{\normalfont f, Y}}$), we begin by calculating TDOS $\rho_0(\omega)$ at the junction site $x=0$, for $V=1, 0$ and $-1$, as a function of frequency $ \omega$, as shown in Fig.~\ref{fig2}. We notice that the TDOS of junction site near the Fermi-energy for $V=-1$ shows a peak at $\omega \rightarrow 0$ which is a signature of enhancement, whereas it shows a suppression near $\omega \rightarrow 0$ for the $V= +1$. {The peak near $\omega \approx 0$ for $V\le 0$ owes its origin to the degeneracy at the Fermi-point of the half-filled fermionic Y junction system, and the TDOS shows Lorentzian behavior with $\omega$ for $\omega \lesssim \eta$, due to the introduction of the broadening factor $\eta$, as discussed for Eq. \eqref{eq.xTDOS} in Sec. \ref{MODEL AND NUMERICAL TECHNIQUE}.} For $V=+1$, the TDOS shows a peak at a large $\omega$, which though similar to 1D SF model, differs in terms of the power law exponent. 
For the 1D SF chain, $\alpha_{\text{\normalfont bulk}}$ and $\alpha_{\text{\normalfont end}}$ are calculated from Eq. \eqref{eqn.powerlaw1D} as $0.04$ and $1/3$, respectively, for $V=+1$. Whereas we extract $\alpha_{\text{\normalfont f, Y}}=0.120\pm0.04$ from the TDOS spectra of the fermionic Y junction system for $V=+1$. For $V=-1$, $\alpha_{\text{\normalfont bulk}}$ and $\alpha_{\text{\normalfont end}}$ are calculated from Eq. \eqref{eqn.powerlaw1D} as $0.08$ and $-1/3$, respectively, but we extract $\alpha_{\text{\normalfont f, Y}}=-1.50\pm0.07$ for the fermionic Y junction. For $V=0$, we find $\alpha_{\text{\normalfont f, Y}}=-0.13\pm0.04$.
On increasing $V$, we notice transition in the nature of TDOS from enhancement to suppression. The repulsive interaction fermionic Y junction model ($V> 0$) shows suppression, whereas in the attractive regime ($V< 0$) it shows enhancement, as reflected by the change in sign of $\alpha_{\text{\normalfont f, Y}}$ in the inset of Fig.~\ref{fig2}. {In the inset of Fig.~\ref{fig2}, the error in extraction of $\alpha_{\text{\normalfont f, Y}}$ has been determined by keeping the lower bound for fitting as $\omega \approx \eta$, and by varying the upper bound of $\omega \in (2\eta,1)$.}

\label{Local Density of States of the bosonic Y junction (BY) Model }

Similar to the fermionic Y junction model, the bosonic Y junction also shows a qualitatively similar TDOS pattern. The TDOS $\rho_0(\omega)$ of the junction site for the bosonic model for various values of $J^z$ are shown as a function of frequency $\omega$ in Fig.~\ref{fig3}.  We notice that TDOS of junction sites $\rho_0(\omega)$ for $J^z=-1/2$ shows enhancement and the corresponding power law exponent is extracted as $\alpha_{\text{\normalfont b, Y}}=-0.69 \pm 0.06$. The maxima of $\rho_0(\omega)$ decreases with increasing $J^z$, and $\rho_0(\omega)$ follows a power law with exponent $\alpha_{\text{\normalfont b, Y}}$ which increases with increasing $J^z$, as shown in the inset of Fig.~\ref{fig3}. At $J^z=+1$ the power law corresponding to the suppression is given by $\alpha_{\text{\normalfont b, Y}}=0.10 \pm 0.04$. {Similar to the fermionic Y junction case, the error in extraction of $\alpha_{\text{\normalfont b, Y}}$ in the inset of Fig.~\ref{fig3} has been determined by keeping the lower bound for fitting as $\omega \approx \eta$, and by varying the upper bound of $\omega \in (2\eta,1)$.}          

While the regime of enhancement and suppression is qualitatively similar for the bosonic and the fermionic Y junctions, especially in the regime of attractive interactions ($J^z<0$ and $V<0$), the quantitative details differ, e.g., in terms of the power law exponent $\alpha$. The power law exponents are $\alpha_{\text{\normalfont b, Y}}= -0.86 \pm 0.08$ for the bosonic Y junction, and $\alpha_{\text{\normalfont f, Y}}= -1.50\pm0.07$ for the fermionic Y junction at $J^z= -1$ and $V=-1$, respectively. 
As discussed before in Sec. \ref{MODEL AND NUMERICAL TECHNIQUE}, this difference could be attributed to the difference in the exchange statistics of the particles of the respective models.

\subsection{Conductance $G_{\beta,\gamma}$ and $M$ Fixed Point}  \label{sub.Conductace}

Since we observed qualitatively similar TDOS enhancements at the junction in both the bosonic and the fermionic Y junction systems, though the TDOS power law exponents differed quantitatively for the two models, it becomes important to identify the stable fixed point the Y junction flows into, to correctly characterize the system. In absence of any external field, the Y junction preserves the time reversal symmetry and can be described by the elusive $M$ fixed point reported in literature.~\cite{2006_Oshikawa} {The $M$ fixed point describes the stable fixed point for the Y junction with the following properties: (1) It must be time reversal invariant, (2) It must be a wire-symmetric junction (symmetric under permutation of the three wires forming the junction), and (3) The bulk Luttinger parameter g should be bounded by $1< g< 3$. 

\begin{figure}[t]
\includegraphics[width=\columnwidth]{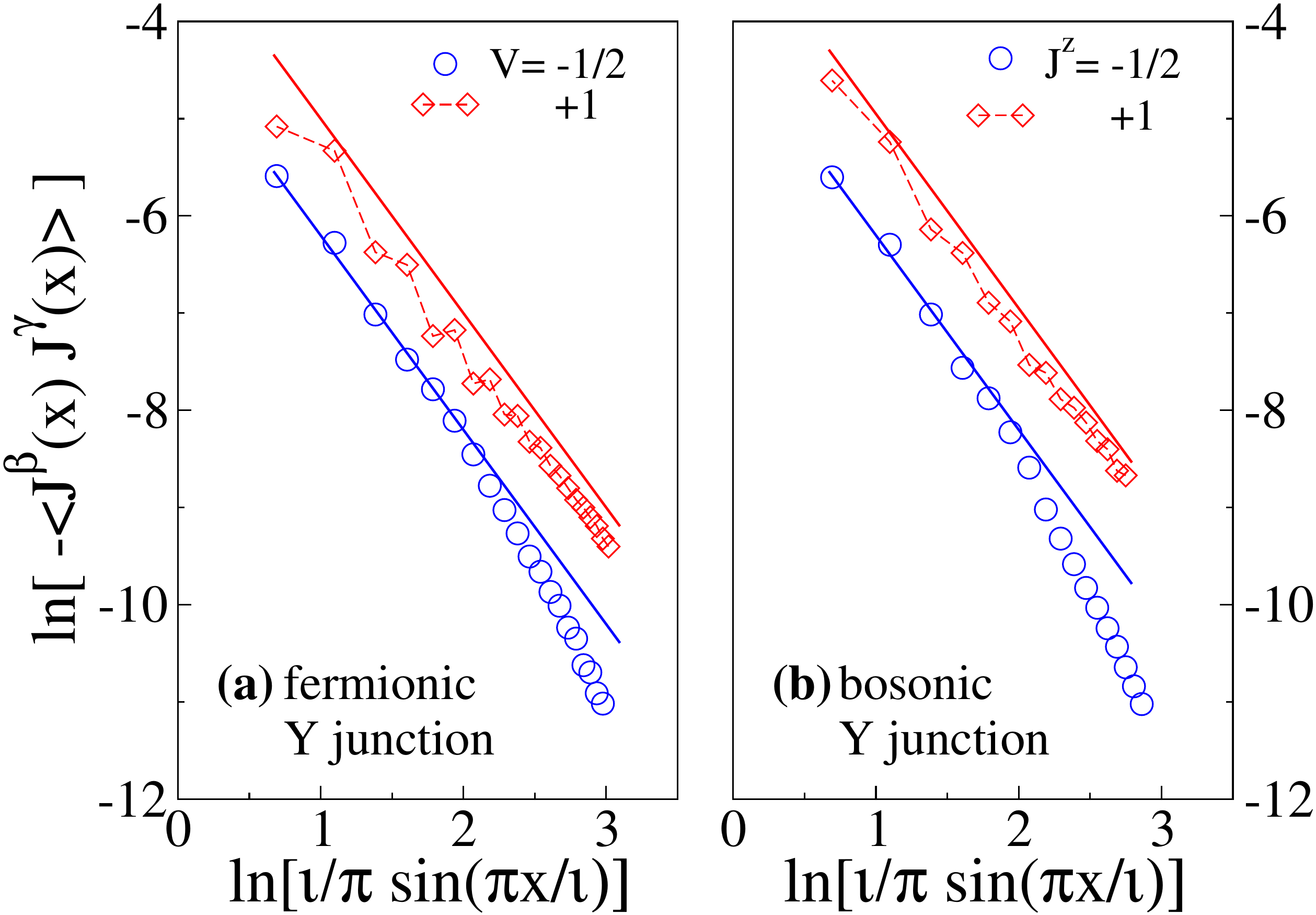}
\caption{(Color online) Log-log plot of the static current-current correlations for the (a) fermionic Y junction system at $V=-1/2$ and $+1$ (or equivalently, $g_f=1.192$ and $3/4$ from Eq. \eqref{eqn.LuttingerSF}), and (b) bosonic Y junction system, at $J^z=-1/2$ and $+1$ (or equivalently, $g_s=3/2$ and $1/2$ from Eq. \eqref{eqn.LuttingerXXZ}), where $x$ represents the site index as shown in schematic Fig.~\ref{fig1: Yjn}, and $\ell=(N-1)/3$ represents the length of each arm forming the Y junction. Here, $N=310$. Solid lines are of slope $=-2$.} \label{fig4}
\end{figure} 

It is well known that the bosonization description of the $M$ fixed point is not possible, and that only the numerical study of the same can be conducted.~\cite{2006_Oshikawa,2012_Rahmani}} Rahmani et al. investigated a fermionic Y junction model with periodic boundary conditions at half-filling where they developed a boundary conformal field theory based approach to find the DC conductance in these systems.~\cite{2012_Rahmani} 
At the $M$ fixed point of this Y junction in the regime of attractive interactions ($1<g<3$), the following relation is expected to be followed away from the boundary, i.e., for $\ell \rightarrow \infty$ and $x \rightarrow \infty$~\cite{2012_Rahmani}:
\begin{equation} \label{eqn.Conductance}
 G_{\beta \gamma} = \underset{ x \rightarrow \infty}{lim} \langle J^{\beta}_{R}(x)J^{\gamma}_{L}(x) \rangle_{gs} \left[ 4 \ell sin\left( \dfrac{\pi}{\ell} x \right) \right]^{2} \dfrac{e^2}{h} ,
\end{equation}
where $J^{\beta}_{R}(x)$ and $J^{\gamma}_{L}(x)$ represent the right-moving and left-moving chiral current on any constituent wire $\beta$ and $\gamma$ of the Y junction, respectively, and $\ell$ is the length of each arm of the Y junction system. In the fermionic Y junction model, the current is simply given by $J(x) = i \left( c^{\dagger}_{x}c_{x+1}-c^{\dagger}_{x+1}c_{x}\right)$, where $c^{\dagger}_{x} (c_{x})$ represents the creation (annihilation) operator acting at site $x$. Similarly for spin system,  $J (x) = i \left( S^{+}_{x}S^{-}_{x+1}-S^{+}_{x+1}S^{-}_{x} \right)$, where $S^{+}_x(S^{-}_x)$ are the spin raising (lowering) operators acting at site $x$. In the finite $x/\ell$ limit, for $\ell \rightarrow \infty$ and $x \rightarrow \infty$, $G_{\beta \gamma}$ should have a constant value and the following relation is expected to hold:
\begin{equation} \label{eq.Current-Current}
\langle J^{\beta}(x)J^{\gamma}(x) \rangle_{gs} \propto \left[ \dfrac{1}{\pi / \ell} sin\left( \dfrac{\pi}{\ell} x \right) \right]^{-2}
\end{equation}

We plot $\langle J^{\beta}(x)J^{\gamma}(x) \rangle_{gs}$ as a function of $\left[ \dfrac{1}{\pi / \ell} sin\left( \dfrac{\pi}{\ell} x \right) \right]$ in log-log scale in Fig.~\ref{fig4} to confirm the validity of this relation. Figs.~\ref{fig4}(a) and \ref{fig4}(b) correspond to the fermionic and bosonic Y junction models, respectively. We observe that in both the attractive limit $V<0$ $(J_z<0)$ and the repulsive limit $V>0$ $(J_z>0)$, the slope is found to be in the vicinity of $-2$ (represented by solid lines in Fig.~\ref{fig4}(a) and Fig.~\ref{fig4}(b)), which is consistent with previous works.~\cite{2012_Rahmani} The oscillatory nature of the static current-current correlations is clearly visible in the repulsive $V>0$ ($J^z>0$) limit, again consistent with Ref.~[\onlinecite{2012_Rahmani}]. This strongly suggests that our Y junction systems could be in the vicinity of the $M$ fixed point, in the attractive regime of interaction $V<0$ ($J^z<0$), which is also the same interaction regime where we report the enhancement in TDOS of the junction in Sec.~{\ref{sub.TDOS}}. {Since the prediction of the existence of $M$ fixed point~\cite{2006_Oshikawa}, not much was known about it except for its existence, until the DC conductance related to this fixed point was reported in Ref.~[\onlinecite{2012_Rahmani}]. Even then, the dynamical properties and power law exponents related to this fixed point remained unknown until now. In the present work we show the relation of a stable $M$ fixed point with the enhancement of the TDOS in the attractive regime of interactions, and thus contribute new information regarding this fixed point to the literature of multi-wire junctions. We note here that both the bosonic and the fermionic Y junctions follow Eq. \eqref{eq.Current-Current} in the attractive interaction regime ($V<0$ and $J^z<0$, respectively), although the respective power laws for the TDOS enhancement are different, as discussed in Sec. \ref{sub.TDOS}.}

\subsection{Length Scale of TDOS Enhancement}  \label{sub.Cutoff}

\begin{figure}
\centering
\includegraphics[width=\columnwidth]{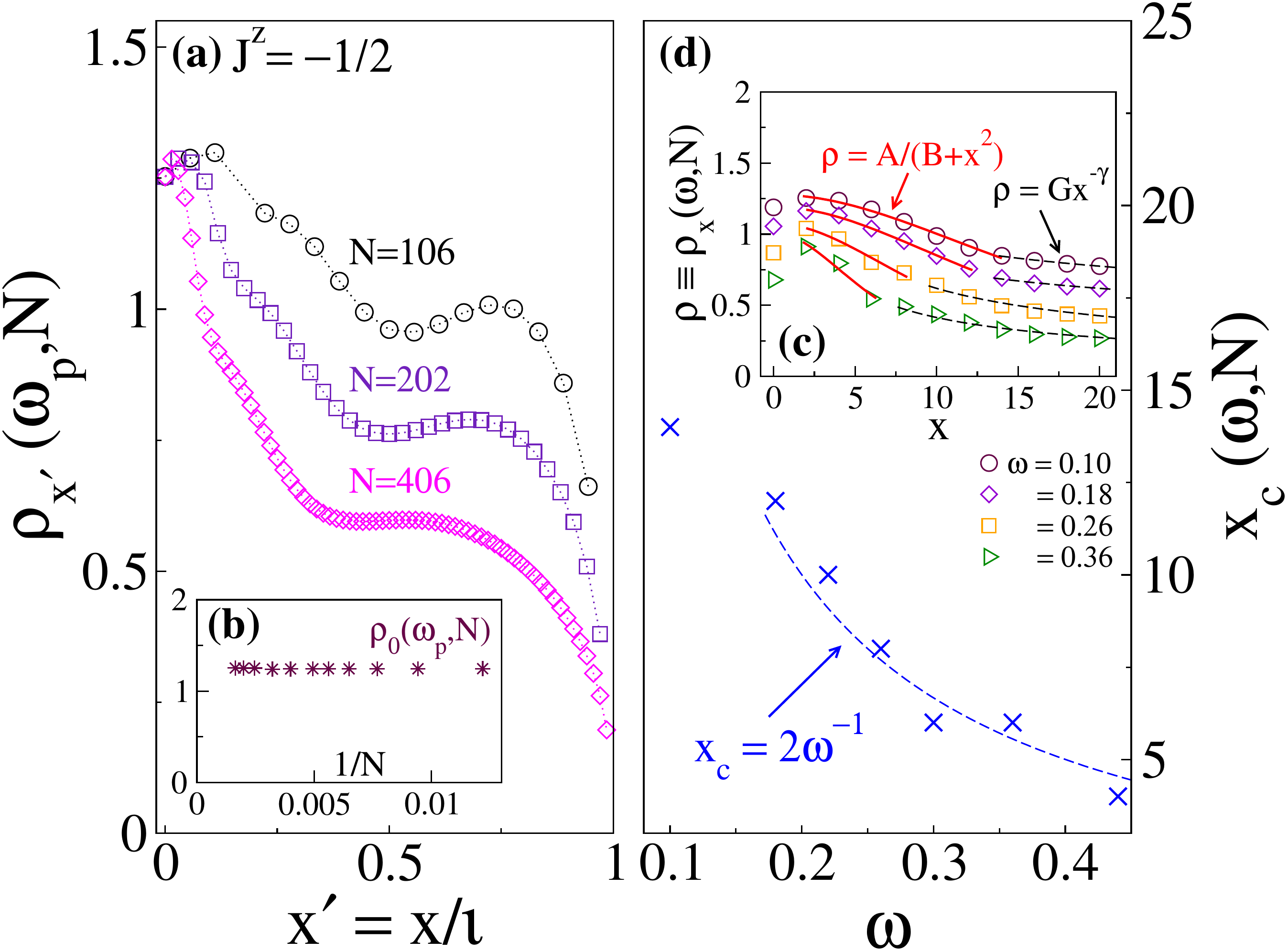} 
\caption{(Color online) (a) Plot of TDOS $\rho_{x^{\prime}} (\omega_p, N)$ vs. $x^{\prime}$, for different $N$ at $J^z=-1/2$ (or equivalently, $g_s=3/2$ from Eq. \eqref{eqn.LuttingerXXZ}), for the bosonic Y junction, where, $x^{\prime}=x/\ell$, and $\ell=(N-1)/3$ is the length of each constituent chain, and $x^{\prime}=0$ represents the junction site (ref. to schematic in Fig.~\ref{fig1: Yjn}). $\omega_p$ is the peak frequency where the maxima of the TDOS spectra occurs for a particular $J^z$ and $N$. ($J=1$ is kept fixed). (b) Plot of TDOS maxima at the junction $\rho_{0} (\omega_p, N)$ vs. inverse system size $1/N$, for the same parameters as in Fig.(a). (c) Plot of TDOS $\rho_{x} (\omega, N)$ vs. $x$, for different $\omega$ at $N=406$ and $J^z=-1/2$, for the bosonic Y junction. For $x<x_c$, the fitting is Lorentzian: $\rho_{x} (\omega, N) = A/(B+x^2)$, and is represented by the solid red curves. The parameters $(A,B)$ extracted for $\omega=0.10,0.18,0.26,0.36$ are $(465.24,363.90)$, $(298.05,250.25)$, $(133.48,124.078)$, and $(47.69,47.26)$, respectively. For $x \ge x_c$, the fitting follows a power law: $\rho_{x} (\omega, N) = Gx^{-\gamma}$, and is represented by the dashed black curves. The parameters $(G,\gamma)$ extracted for $\omega=0.10,0.18,0.26,0.36$ are $(1.60,0.24)$, $(1.44,0.28)$, $(2.14,0.54)$, and $(1.64,0.60)$, respectively. (d) Plot of distance from the junction up to which the Lorentzian fitting holds, $x_c$, as a function of $\omega$, for the same parameters as in Fig.(c). It can be fitted with power law of the form, $x_c={2}/{\omega}$. } \label{fig5}
\end{figure}

So far, we have illustrated that the bosonic and the fermionic Y junctions are connected to a stable $M$ fixed point in the parameter regime $1<g<3$ or the attractive interaction limit ($V<0$ or $J^z<0$). Existing studies in literature regarding the effect of impurities in quantum wires point to a finite spatial cut-off on the enhancement caused by the impurities~\cite{2006_Kakashvili}. In similar spirit, we wish to study the spatial extent of the enhancement observed in the attractive limit of the Y junction. Since the TDOS spectra of the bosonic and the fermionic model on Y junctions are similar, here we present the results of only the bosonic Y junction model. To estimate the spatial extent of the enhancement in TDOS, we plot the maximum intensity of TDOS $\rho_{x^{\prime}}(\omega_p)$ at peak frequency $\omega_p$ as a function of scaled distance from the junction $x^{\prime}=x/\ell$  in Fig.~\ref{fig5} for the bosonic Y junction. $\rho_{x^{\prime}}({\omega}_p)$ is inversely proportional to $\eta$ in case of resonance condition and proportional to the sum of the squares of all the transition  matrix elements $\frac{1}{\eta}\sum_{n} | \langle \psi_n | S_{x}^{+} | \psi_{0} \rangle |^{2}$. Therefore, keeping the $\eta$ same, we can extrapolate the sum of the matrix elements for different $N$. The spatial dependence of $\rho_{x^{\prime}}(\omega_p)$ as function of scaled spatial unit $x^{\prime}$ at $J^{z}=-1/2$ (enhancement regime) for three system sizes $N=106, 202$ and $406$ are shown in Fig.~\ref{fig5}(a). The finite size dependence of $\rho_{x^{\prime}}(\omega_p)$ for sites near the junction is weak as shown in Fig.~\ref{fig5}(b), but it is strong for sites away from the junction, as is clear from Fig.~\ref{fig5}(a). We also note that the extent of enhancement of $\rho_{x^{\prime}}(\omega_p)$ is limited to the neighborhood of the junction. 

\label{sub.LengthScale}
{To study the length scale of enhancement near the junction in more detail, in Fig.~\ref{fig5}(c) we plot the TDOS $\rho_x(\omega)$ as function of $x$, for different frequencies $\omega$, for $N=406$ and $J^z=-1/2$. Near the junction, the TDOS follows a Lorentzian behavior of the form $A/(B+x^2)$ with $x$ for $0<x<x_c$, whereas it follow an algebraic decay of the form $Gx^{-\gamma}$ for $x>x_c$. We recognize the distance which shows this transition from Lorentzian fitting to power law fitting, $x_c$, as the length scale of TDOS enhancement. We note that $x_c$ decreases continuously with $\omega$ and eventually tends towards $x_c \approx 3 \pm 1$, as evident from the shrinking Lorentzian fitting regime of $\rho_x(\omega,N)$ with $x$ in Fig.~\ref{fig5}(c), and shown more clearly in Fig.~\ref{fig5}(d). This result is consistent with an earlier prediction for bosonizable fixed points of the Y junction which predicts a relation between the length scale of enhancement of TDOS and the frequency scale of tunneling $\omega$ as, $x_c \propto 1/\omega$.~\cite{2009_Agarwal} From our analysis we conclude that for the symmetrically coupled Y junction, at $M$ fixed point, the enhancement of TDOS is highly localized near the junction for moderate values of $\omega$.}

\section{Summary and Conclusion} \label{SUMMARY}
{Junction of TLL wires poses a complex quantum impurity problem owing to the richness of the manifold of fixed point that it can host. In this paper, we have considered the simplest possible Y junction comprising of three equi-length 1D TLL wires which are symmetrically coupled to the central junction site. Using dynamical DMRG, we have calculated TDOS as a function of distance from the junction and extracted the associated power law exponents. We observe enhancement in TDOS in the attractive interaction limit $(1 < g < 3)$, and suppression of TDOS in the repulsive interaction limit $(g < 1)$ in case of both the bosonic Y junction and the fermionic Y junction, though they follow distinct power law exponents for the TDOS. This difference can be attributed to the non-trivial many-body phase factors associated in the hopping between the junction site and the constituent arms, and stems from the different quantum exchange statistics of constituent particles. Earlier Oshikawa et al.~\cite{2006_Oshikawa} had conjectured the existence of a ``mysterious" stable $M$ fixed point for such a system in the regime of $1 < g < 3$, however its properties had remained unknown as this fixed point is not bosonizable. Later on, Rahmani et al.~\cite{2012_Rahmani} evaluated the static ground state correlation function for the $M$ fixed point using time-independent DMRG. In this work we perform a numerical analysis which is complimentary to Rahmani et al. where we use the dynamical DMRG to evaluate the dynamical correlation functions. These dynamical correlation functions are then used to evaluate the TDOS for the $M$ fixed point. 

As far as a quantitative comparison with exsiting bosonization prediction is concerned, one could compare the power law that is numerically obtained in our work for the $M$ fixed point with the existing prediction of power laws for all possible bosonizable fixed points which respect time reversal symmetry and wire symmetry (symmetric under permutations of the three wires among themselves), as these two symmetries are valid symmetries for our numerical analysis. We find that, if we try to extract the parameter $\theta$ (the parameter parametrising the space of fixed points respecting these symmetries) from Eq. (6) of Ref.~[\onlinecite{2009_Agarwal}], which describes the power law exponents for these bosonizable fixed points, it gives unphysical solution leading to the condition of $cos \theta > 1$, for the attractive regime of interaction $1 < g < 3$. This can be considered as an illustration of the fact that the $M$ fixed point cannot be described through bosonization analysis. 

Finally, we investigated the spatial extent of TDOS enhancement through a finite size scaling study and observed that the TDOS peak amplitude near the junction is weakly dependent on the system size $N$. We also checked that the length scale of enhancement showed a $1/\omega$ dependence on the frequency, which is consistent with an earlier study that reported enhancement of the Y junction TDOS for various bosonizable fixed point studies therein~\cite{2009_Agarwal}. We noted that for $\omega > \eta$ the TDOS enhancement spans over just a few sites away from the junction, e.g., $3 \pm 1$ lattice units for $N = 406$ and $J^z = -1/2$. Thus, we found that the enhancement of the TDOS is highly localized near the junction site.}

\begin{acknowledgments} \label{Acknowledgement}
{M.S.R. thanks S. N. Bose National Centre for Basic Sciences for PBIR-PhD fellowship. M.K. thanks D. Sen, S. Ramasesha, and Z. G. Soos for valuable suggestions. M.K. thanks Department of Science and Technology (DST), India for Ramanujan fellowship and computation facility provided under the DST Project No. SNB/MK/14-15/137. S.D. would like to acknowledge the ARF grant received from IISER Kolkata and the MATRICS grant(MTR/ 2019/001 043) from Science and Engineering Research Board (SERB), India for funding.} 
\end{acknowledgments}

\label{References}

%

\end{document}